\documentclass{aastex62}

\usepackage{longtable}

\graphicspath{{./}{figures/}}

\accepted{22 January 2021}
\submitjournal{ApJ}

\shorttitle{First Results on Yellowballs from DR2}
\shortauthors{Wolf-Chase et al.}

\begin{document}

\title{The Milky Way Project: Probing Star Formation with First Results on Yellowballs from DR2}

\correspondingauthor{Grace Wolf-Chase}
\email{gwchase@psi.edu}

\author[0000-0002-9896-331X]{Grace Wolf-Chase}
\affil{Planetary Science Institute \\
1700 East Fort Lowell, Suite 106 \\
Tucson, AZ 85719 USA}
\affil{The Adler Planetarium \\
Astronomy Department \\
1300 South Lake Shore Drive \\
Chicago, IL 60605, USA}

\nocollaboration

\author[0000-0003-1539-3321]{C. R. Kerton}
\affiliation{Iowa State University\\
Department of Physics and Astronomy\\
2323 Osborn Dr.  \\
Ames, IA 50011, USA}
\nocollaboration

\author[0000-0002-3723-6362]{Kathryn Devine}
\affiliation{The College of Idaho\\
2112 Cleveland Blvd \\
Caldwell, ID 83605, USA}
\nocollaboration

\author{Anupa Pouydal}
\affiliation{The College of Idaho\\
2112 Cleveland Blvd \\
Caldwell, ID 83605, USA}
\nocollaboration

\author{Johanna Mori}
\affiliation{The College of Idaho\\
2112 Cleveland Blvd \\
Caldwell, ID 83605, USA}
\nocollaboration

\author{Leonardo Trujillo}
\affiliation{The College of Idaho\\
2112 Cleveland Blvd \\
Caldwell, ID 83605, USA}
\nocollaboration

\author{Aurora Cossairt}
\affiliation{The College of Idaho\\
2112 Cleveland Blvd \\
Caldwell, ID 83605, USA}
\nocollaboration

\author{Sarah Schoultz}
\affiliation{The College of Idaho\\
2112 Cleveland Blvd \\
Caldwell, ID 83605, USA}
\nocollaboration

\author[0000-0002-6244-477X]{Tharindu Jayasinghe}
\affiliation{The Ohio State University\\
Department of Astronomy\\
140 West 18th Ave \\
Columbus, OH 43210, USA}
\affiliation{The Ohio State University\\
Center for Cosmology and Astroparticle Physics\\
191 W. Woodruff Ave \\
Columbus, OH 43210, USA}
\nocollaboration

\author[0000-0001-9062-3583]{Matthew Povich}
\affiliation{California State Polytechnic University \\
Department of Physics and Astronomy \\
3801 W. Temple Ave \\
Pomona, CA 91768, USA}
\nocollaboration

\begin{abstract}
Yellowballs (YBs) were first discovered during the Milky Way Project citizen-science initiative (MWP; \citealt{sim12}). MWP users noticed compact, yellow regions in \emph{Spitzer Space Telescope} mid-infrared (MIR) images of the Milky Way plane and asked professional astronomers to explain these “yellow balls.” Follow-up work by \citet{ker15} determined that YBs likely trace compact photo-dissociation regions associated with massive and intermediate-mass star formation. YBs were included as target objects in a version of the Milky Way Project launched in 2016 \citep{jay19}, which produced a listing of over 6000 YB locations. We have measured distances, cross-match associations, physical properties, and MIR colors of $\sim 500$ YBs within a pilot region covering the $ l=30^{\circ}-40^{\circ}, b=\pm 1^{\circ}$ region of the Galactic plane. We find $\sim 20-30\%$ of YBs in our pilot region contain high-mass star formation capable of becoming expanding \ion{H}{2} regions that produce MIR bubbles. A majority of YBs represent intermediate-mass star-forming regions whose placement in evolutionary diagrams suggest they are still actively accreting, and may be precursors to optically-revealed Herbig Ae/Be nebulae. Many of these intermediate-mass YBs were missed by surveys of massive star-formation tracers and thus this catalog provides information for many new sites of star formation. Future work will expand this pilot region analysis to the entire YB catalog.

\end{abstract}


\section{Introduction} \label{sec:intro}

Yellowballs (YBs), named for their appearance in \emph{Spitzer Space Telescope} 4.5-8.0-24 $\mu$m mid-infrared (MIR) images, were originally discovered by citizen scientists working on the Milky Way Project \citep[MWP;][]{sim12}. The only study of YBs to date, \citet{ker15} (KWA15 hereafter), attempted to answer the citizen scientists' question, ``What are these yellow balls?" by showing that YBs likely represent young photo-dissociation regions (PDRs) associated with intermediate and massive star-forming regions. They proposed a sequence in which some fraction of YBs will later expand their PDRs and become MIR bubbles. 

KWA15 cataloged 928 YBs identified by MWP participants and used infrared photometry, together with cross-matching YBs with various catalogs, to explore YB properties. Since YBs were a serendipitous discovery, the original MWP tools were not designed with them in mind, and the MWP tutorial materials did not instruct users to identify YBs or provide them with interactive tools to accurately measure YB angular sizes. 

Although sizes were included in the KWA15 study, at best these were rough upper limits to the actual angular sizes of the YBs. Also, since users identified these objects on an ad-hoc basis, there was no expectation the resulting catalog was complete.  Cross-matches of YBs with catalogs produced by the Bolocam Galactic Plane Survey \citep[BGPS;][]{agu11} and ATLASGAL \citep{cse14} indicated that nearly all of the KWA15 YBs are associated with dense gas. Additionally, $\sim$ 65\% of KWA15 YBs have counterparts in the {\it WISE} catalog of Galactic \ion{H}{2} regions \citep{and14} and $\sim$ 34\% in the Red MSX Source (RMS) catalog \citep{lum13}, to within a 24\arcsec cross-match distance.  Most of the YBs reported in KWA15 appear in environments associated with star formation, such as infrared dark clouds (IRDCs) and/or in close proximity to the bubbles representing expanding \ion{H}{2} regions \citep{sim12}. 

KWA15's IR photometry indicated many YBs have higher flux densities at 8 $\mu$m than 12 $\mu$m, with flux densities rising again toward longer IR wavelengths. The brightness of these objects at 8 $\mu$m could be explained if YBs are dense, young PDRs containing a high fraction of ionized polycyclic aromatic hydrocarbon (PAH) emission \citep[e.g.,][]{roe96}. This could also explain why many YBs do not have counterparts in the RMS catalog \citep{lum13}, since having rising SEDs across MIR MSX wavelengths was employed as one criterion for inclusion in the RMS catalog. For a sample of 138 YBs that had counterparts in the RMS catalog and good distance estimates, KWA15 found that these YBs had typical diameters of 0.1 -- 1 pc and luminosities of $10^3$ -- $10^6$ L$_\sun$. However, this sample was likely biased to luminous objects, since the RMS survey targeted the identification of massive YSOs, complete to typical B0 stars at the distance of the Galactic center \citep{lum13}.  

Based on their findings, KWA15 argued that YBs represent young PDRs associated with intermediate and massive star-forming regions. However, KWA15's results raised new important questions about the nature of YBs. Does the RMS-selected sample from KWA15 accurately represent the YB luminosity and mass distribution? What are the properties of the protostellar clumps associated with YBs? What fraction of the YBs contain the massive stars that will eventually form mature \ion{H}{2} regions and large  MIR bubbles? 

To address these questions, a more intentional MWP study was developed to focus on YBs. The user interface was redesigned to include a tool that both identified YBs and measured their sizes. The second data release (MWP DR2) also includes a new catalog of bubbles and bow shocks driven by OB stars \citep{jay19}. This paper focuses on a pilot study utilizing a 20 square degree subset of the DR2 YBs. Analysis of the complete DR2 YB catalog will be addressed in a subsequent paper. In addition to resolving questions about the nature of YBs, the YB catalog will provide a new, extensive catalog containing the properties of thousands of star-forming regions. Many of these regions, particularly in the intermediate-mass range, have not been studied or cataloged before. 

The current list of DR2 YBs and our motivation for analyzing a subset of these objects prior to addressing the entire survey are discussed in \autoref{sec:catalog}. A comparison of DR2 YBs with existing catalogs tracing star-forming regions and dense gas structures is described in \autoref{sec:cats}. Our derivation of YB distances is presented in \autoref{sec:distances}, and the physical properties of interstellar medium clumps associated with YBs are summarized in \autoref{sec:props}. Our photometry procedure is described in \autoref{sec:photo}, and results of a 2D-Gaussian fitting routine we employed to assess the accuracy of user-measured sizes and quantify the deviation of features from those expected for young PDRs are presented in \autoref{sec:fits}. Our analysis and discussion of YB properties are presented in \autoref{sec:analysis}.  

Our conclusions are presented in \autoref{sec:conclusions}. 

\section{A First Look at Yellowballs in Milky Way Project DR2} \label{sec:catalog}

The original MWP launched in 2010 as the ninth online research program hosted through the Zooniverse suite of citizen science programs \citep{sim14}. The MWP was created to enable large statistical studies of Galactic star formation using MIR data from the \emph{Spitzer} GLIMPSE \citep{ben03} and MIPSGAL \citep{car09} surveys. The original release of the MWP focused on identifying and measuring MIR bubbles \citep{sim12}. In this original data analysis, YBs were identified by users as interesting but unexplained objects. Users flagged these objects for examination by astronomers. Follow-up investigation resulted in the work presented in KWA15.

In September 2016, an implementation of the MWP was released to citizen scientists that included the identification and measurement of YBs and bow shocks produced by winds from OB stars as two additional key science objectives. Citizen scientists were trained to mark the center location and angular size of YBs in \emph{Spitzer} 4.5-8.0-24 $\micron$ images.

Employing this new tool, users searched for YBs in images from the \emph{Spitzer} GLIMPSE and MIPSGAL legacy surveys of the inner Galactic plane ($|l|<65^{\circ}$); the Cygnus-X legacy survey, an approximately 24 square degree region (centered on $ l\sim 79\fdg3, b\sim 1^{\circ}$) of one of the richest known star-forming regions in our Galaxy \citep{hor07,bee10}; and the SMOG legacy survey \citep{car08}, which covers a 21 square degree ($l=102^{\circ}$ to $109^{\circ}$, $b=0^{\circ}$ to $3^{\circ}$) area of a representative region of the outer Galaxy. 

The MWP history, interface, images, workflow, and identification tools, as well as the construction of the Milky Way Project DR2 bubble and bow shock catalogs, are presented in detail in \citet{jay19}. Here, we summarize specifics for producing the current list of YBs. Citizen-scientist classifications were aggregated to produce individual catalog entries for any given object. The MWP data reduction pipeline used the density-based clustering algorithm DBSCAN \citep{est96} with a 0.002 degree clustering radius to identify YBs, identical to the process used to produce the DR1 small bubble catalog \citep{sim12}. In this pipeline, a minimum of five classifications were needed to identify a YB cluster. For each cluster, the Galactic longitude (l), Galactic latitude (b), and YB radius (r) were averaged. The uncertainties in these parameters were calculated based on their standard deviations among the classifications in each cluster. The hit rate for a given cluster is the ratio of the total number of classifications in the cluster to the total number of times images containing the classified object were viewed by MWP users. DR2 YBs were identified from images at the highest zoom level only ($0.15^{\circ} \times 0.075^{\circ}$). 

The complete list of DR2 YBs is presented in \autoref{tab:dr2ybcat}. It contains 6,176 YBs, more than six times the number of YBs investigated by KWA15, and more than twice the number of bubbles in the DR2 catalog. It includes YB position in Galactic coordinates, the user-measured YB radius in degrees, dispersion in YB position in degrees, dispersion in YB radius in degrees, and hit rate. This list includes all identified YBs. The lowest reported hit rate is 0.05; however, 6,050 YBs have hit rates $>$ 0.125, the cutoff employed for DR2 bubbles. We discuss the relationship between different measurements used to estimate YB sizes in \autoref{sec:size}. We will refer to the user-defined radius as the MWP radius hereafter. No filtering has yet been applied to remove spurious objects and produce a final version of the YB catalog. Instead, in this paper we present a pilot study that uses all YBs identified by citizen scientists in the $ l=30^{\circ}-40^{\circ}, b=\pm 1^{\circ}$ region to develop methods for assessing the reliability of citizen-scientist identifications and automating procedures to analyze the entire YB catalog.

\begin{deluxetable}{lccccccc}
\tablecaption{DR2 Yellowballs \label{tab:dr2ybcat}}
\tablewidth{0pt}
\tabletypesize{\normalsize}
\tablehead{\colhead{YB}  & \colhead{Gal. Longitude} & \colhead{Gal. Latitude}       & \colhead{MWP Radius}    & \colhead{$\sigma_l$} & \colhead{$\sigma_b$} & \colhead{$\sigma_r$} & \colhead{Hit} \\
        \colhead{Number} & \colhead{(degrees)}      & \colhead{(degrees)}           & \colhead{(degrees)} & \colhead{(degrees)} & \colhead{(degrees)}   & \colhead{(degrees)}  & \colhead{Rate}
}
\startdata
1 & -0.04143 &  0.16705 & 0.00366 & 0.00077 & 0.00073 & 0.0020  & 0.65 \\
2 & -0.02311 &  0.16916 & 0.00391 & 0.00056 & 0.00053 & 0.0014  & 0.48  \\
3 & 0.04088  &  0.02020 & 0.00314 & 0.00019 & 0.00037 & 0.00094 & 0.23 \\
4 & 0.18604  & -0.61181 & 0.00304 & 0.00071 & 0.00036 & 0.0025  & 0.19  \\
5 & 0.20075  & -0.51336 & 0.00222 & 0.00050 & 0.00045 & 0.00090 & 0.29 \\
6 & 0.20950  & -0.00195 & 0.00334 & 0.00039 & 0.0010  & 0.00079 & 0.17 \\
7 & 0.27904  & -0.48463 & 0.00659 & 0.00055 & 0.0010  & 0.0019  & 0.63 \\
8 & 0.31351  & -0.20354 & 0.00456 & 0.00078 & 0.00078 & 0.0016  & 0.53 \\
9 & 0.31413  & -0.19272 & 0.00459 & 0.00079 & 0.0010  & 0.0021  & 0.50 \\
10 & 0.33171 & -0.06248 & 0.00442 & 0.00084 & 0.00033 & 0.0020  & 0.29 \\
\enddata
\tablecomments{This table is available in its entirety in machine-readable format. A portion is shown here for guidance regarding its form and content.}
\end{deluxetable}

\citet{jay19} employed criteria to determine whether bubbles and bow shocks identified by citizen scientists would be included in the DR2 catalogs, and to set reliability flags for objects that were included. These criteria were developed based on prior identifications or knowledge of these objects. In the case of YBs, it is not possible to use the DR1 YB catalog to help assess the reliability of DR2 identifications for two reasons. First, identifications of DR2 YBs were made at four times the resolution of DR1 YBs, which greatly affects the types of objects that might be identified as round and compact in the images. Second, citizen scientists were never specifically asked to search for YBs in DR1, so comparing hit rates between DR1 and DR2 lacks meaning. 

Thus, we have opted to perform an initial analysis of all the YBs that were included in the $ l=30^{\circ}-40^{\circ}, b=\pm 1^{\circ}$ region after clustering and averaging was performed. This region contains 516 YBs, with a negligible 10 YBs that achieved hit rates less than the 0.125 cutoff employed for DR2 bubbles. The region was chosen because it is covered in numerous other Galactic plane surveys and overlaps with multiple catalogs, which are discussed further in \autoref{sec:cats}. 

\section{Results} \label{sec:results}

\subsection{Catalog Cross-Matches} \label{sec:cats}

To help identify those YBs that are most likely associated with regions of star formation we cross-matched the 516 YBs in our pilot region with catalogs of star-forming regions and interstellar medium structures produced from surveys covering this region. We used the Hi-GAL compact source catalogue of \cite{eli17} (EMS17 hereafter) containing 4764 sources within the pilot region, the GaussClump Source Catalogue from the APEX Telescope Large Area Survey of the Galaxy \citep[ATLASGAL;][624 sources]{cse14}, the Co-Ordinated Radio `N' Infrared Survey for High-mass star formation catalog \citep[CORNISH;][494 sources]{pur13}, the Red MSX Source Survey catalog \citep[RMS;][210 sources]{lum13}, and the WISE Catalog of Galactic \ion{H}{2} Regions \citep[][672 sources]{and14}. 

The Hi-GAL and ATLASGAL catalogs both identify compact core and clump structures in the interstellar medium using infrared and submm images respectively. Not surprisingly, 93\% of the ATLASGAL associations are also Hi-GAL associations. The CORNISH, RMS, and the WISE catalogs were all primarily focused on the discovery of regions of higher-mass star formation using radio and infrared emission. CORNISH detected 5-GHz radio emission from \ion{H}{2} regions, and is complete to emission from Galactic sources for B2 and earlier stars.  RMS used infrared colors and fluxes to identify potential high-mass star-forming regions, and is complete to Galactic sources with luminosities corresponding to embedded B0 stars. The WISE catalog was constructed by identifying \ion{H}{2} region candidates via searching for their characteristic MIR morphology.
Exact completeness limits are hard to define for the WISE catalog, but we expect it to include lower luminosity sources that did not meet the detection limits of the CORNISH and RMS catalogs.

The catalogs were obtained from project-specific web sites or from the VizieR catalogue access tool, and the cross-matching was done with TOPCAT \citep{topcat}. A best match symmetric option was chosen to allow for only one-to-one matches and a match radius of 24\arcsec (the average angular diameter of the YBs as defined by the citizen scientists) was used. We found 385 Hi-GAL matches, 162 ATLASGAL matches, 59 CORNISH matches, 88 RMS matches, 263 WISE matches, and 74 objects lacking matches in any of these catalogs. The resulting cross-matches are listed in \autoref{tab:xmatch}. 

To estimate the level of spurious associations, we created twenty simulated catalogs of 516 YBs, by selecting random Galactic longitudes and latitudes from a uniform distribution in longitude and a Gaussian distribution in latitude (similar to the observed YB distribution). For each simulated YB catalog the cross-matching procedure with each real catalog was repeated as described before, and the average and standard deviation of the number of spurious associations was calculated. For the Hi-GAL catalog we found that $4.5\pm0.9$\% ($23\pm5$) of the fake YBs were cross-matched on average, while for the remaining catalogs this value ranged between $0.4\pm0.3$\% ($2\pm1$) and $1.1\pm0.5$\% ($6\pm3$). We conclude that the level of spurious associations is low enough that the catalog cross-matches are a useful way to isolate YBs with additional evidence of star-formation activity.

The hit rates of YBs have a very broad distribution. Considering the full sample of 516 YBs in our pilot region, the mean hit rate is 0.41, with a standard deviation of 0.19. The mean hit rates for YBs matched with the Hi-GAL, RMS, CORNISH, and WISE catalogs are 0.43, 0.44, 0.46, and 0.48 respectively, each with a standard deviation of $\sim$ 0.19, while the mean hit rate for the 74 YBs lacking any catalog counterparts is 0.33 with a standard deviation of 0.16. Given that all but three of the unmatched YBs are associated with molecular gas, it is unclear whether the difference in mean hit rate reflects mis-identified objects, or simply the fact that many of the unmatched YBs have low fluxes. These objects would have been missed by the catalogs biased to massive star-forming regions, and might not be expected to lie near Hi-GAL clump peaks.

\subsection{LSR Velocities and Distances} \label{sec:distances}

Since we expect the majority of YBs to be associated with molecular clouds, we determined LSR velocities of our YB sample using $^{13}$CO emission coincident with YB locations. The $^{13}$CO spectra were obtained from the Boston University--Five College Radio Astronomy Observatory Galactic Ring Survey \citep[BU--GRS;][]{jac06}. We wrote a Python program to automatically identify peaks in the $^{13}$CO spectra at the location of each YB, and isolate emission at a $3 \sigma$ level above the spectrum's baseline, where the noise level ($\sigma$) was determined from emission-free end channels in each spectrum. In cases where there were multiple velocity components, we assumed that the strongest $^{13}$CO emission was associated with the YB. Velocities for 472 YBs were automatically determined this way. Spectra towards the remaining 44 YBs were visually inspected; 21 YB velocities were determined, and 23 YBs were not found to be associated with any significant $^{13}$CO emission. Twenty-six of the YBs in the pilot region were also associated with dense cores observed in NH$_3$ by \citet{wie12}. All of the NH$_3$ velocities were consistent with our measured $^{13}$CO velocities (within 2 km s$^{-1}$), supporting the assumption that the $^{13}$CO peaks are likely associated with molecular clouds hosting YBs. The LSR velocities for the 493 YBs with $^{13}$CO associations are reported in \autoref{tab:ybvel}.

Distances were then calculated using the Bayesian distance calculator developed by \citet{rei16}. This calculator uses $l, b,$ radial velocity and probability of association with the near/far distance as input parameters. The \citet{rei16} distance calculator utilizes a model of the Milky Way spiral arms generated using trigonometric parallaxes of massive star-forming regions \citep{rei14}. A Bayesian approach that considers kinematic distance, displacement from the plane, and proximity to individual parallax sources is then used to determine a distance probability density function for each source. For objects that are clearly associated with spiral arms, such as YBs, the distances obtained using the Bayesian distance calculator are the best possible distance estimates we can derive from an associated velocity. For inner-Galaxy objects, \citet{rei16} report that for spiral-arm objects the calculator's resolution of the near/far distance ambiguity, even when no additional information is provided, is comparable to a more traditional technique using \ion{H}{1} absorption. Distances obtained using the $^{13}$CO velocities and an equal likelihood of a near/far association are shown in \autoref{tab:ybvel}.

\begin{figure}
\plotone{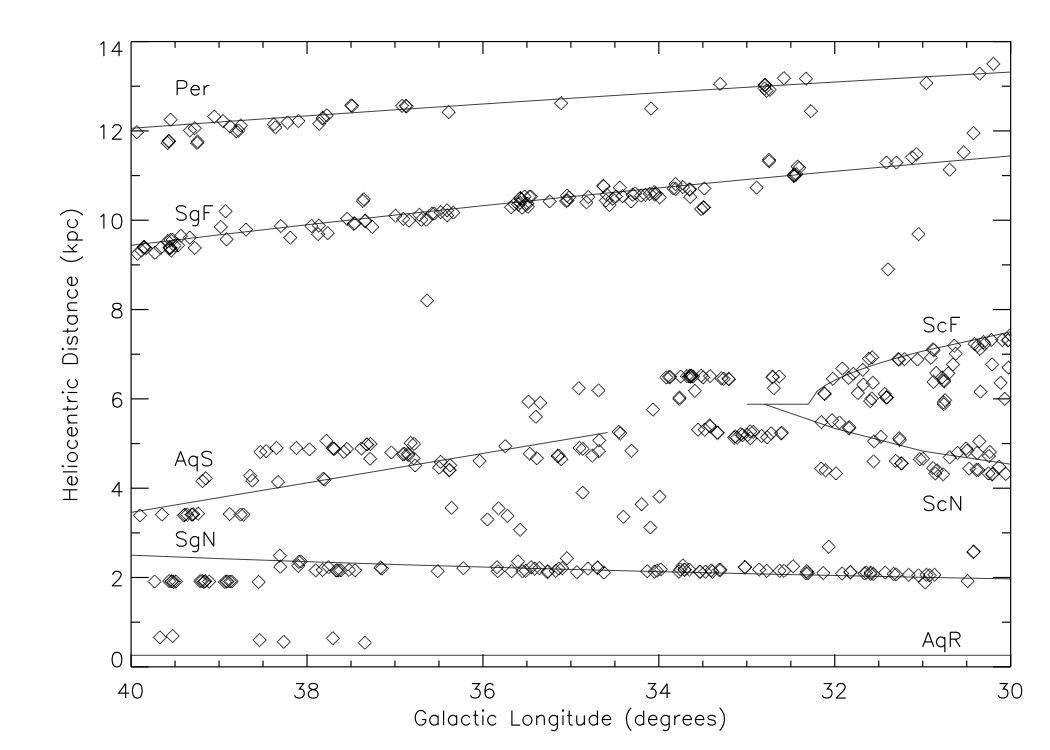}
\caption{YB spatial distribution. Heliocentric distances and Galactic longitudes are plotted for all 493 YBs in the pilot region with CO associations (diamonds). The average statistical uncertainty in the distance is $\sim 0.5$ kpc. Lines show the spiral arm traces along this line of sight from \citet{rei16}: Per = Perseus, SgN and SgF = Sagittarius near and far sections, ScN and ScF = Scutum near and far sections, AqR = the local Aquila Rift feature, and AqS = the Aquila spur between SgN and ScN.}
\label{fig:distances}
\end{figure}

The Hi-GAL compact-source catalog provided additional information about the near/far association for a subset of our YB sample. The Hi-GAL distances were assigned by extracting $^{12}$CO or $^{13}$CO spectra along the line-of-sight to every Hi-GAL source to determine the V$_{LSR}$, and subsequently employing the Galactic rotation model of \citet{bb93}. In the Hi-GAL catalog, three types of flags are used to indicate the quality of the solution to the near/far distance ambiguity (EMS17). An entry of `G' indicates the ambiguity was solved by matching the source position with a catalog of sources with known distances (e.g., \ion{H}{2} regions, masers, etc.) or with features in extinction maps. If none of these was available, the ambiguity was arbitrarily solved in favor of the far distance and a bad-quality (`B') flag was assigned. If a distance estimate was unavailable, a null value (`N') was assigned. When a YB was associated with a Hi-GAL source that had a `G' designation, we assigned this Hi-GAL near/far association to the YB while using the Bayesian distance calculator. For example, if a Hi-GAL source had near distance then we would set P$_\mathrm{far} = 0$. For two of the sources (YB 1166 and 1204) the Bayesian distance calculator and the Hi-GAL near/far solution would not agree due to details of the Galactic model, and we adopted the P$_\mathrm{far} = 0.5$ Bayesian distance. For sources that had a `B' or null flag we used the P$_\mathrm{far} = 0.5$ Bayesian distance. The distribution of our sources in the Galactic plane is shown in \autoref{fig:distances}. As expected, there are noticeable source groupings along the spiral arm tracers (e.g., the SgN arm at $\sim$2 kpc), consistent with the weighting that the \citet{rei16} technique gives to the underlying spiral arm structure model. 

The assigned distances to YBs with Hi-GAL associations are provided in column 4 of \autoref{tab:ybhigal}; for comparison, the kinematic distances reported in the Hi-GAL compact-source catalog are also given in column 6. Variations between our distances and those presented in EMS17 can be attributed to the fact that: 1) the Bayesian model used additional information beyond a kinematic model to calculate distances, and 2) if EMS17 could not resolve the near/far ambiguity, they assumed the far distance, whereas the \citet{rei16} distance calculator uses a Bayesian approach to resolve the ambiguity.

\begin{splitdeluxetable}{lcclllBl llll}
\tablecaption{YB DR2 Pilot Region Cross-matches \label{tab:xmatch}}
\tablewidth{0pt}
\tablehead{\colhead{YB} & \colhead{Gal. Longitude}   & \colhead{Gal. Latitude}   & \colhead{Hi-GAL ID} & \colhead{AGAL ID} & \colhead{CORNISH ID} & \colhead{CORNISH Type} & \colhead{RMS ID} & \colhead{RMS Type} & \colhead{WISE ID} & \colhead{WISE Type} \\
\colhead{Number}     &  \colhead{(degrees)}   &  \colhead{(degrees)}  &          &         &            &              &        &          &         & } 
\startdata
1153 & 30.0033 & -0.2658 & HIGALBM30.0033-0.2654 & G030.0027-0.2643 & \nodata & \nodata & \nodata & \nodata            & G030.003-00.267 &  C      \\
1154 & 30.0233 & -0.0428 & HIGALBM30.0254-0.0389 & \nodata          & \nodata & \nodata & 3110    & Diffuse HII region & G030.022-00.042 &  K      \\
1155 & 30.0242 &  0.1085 & HIGALBM30.0230+0.1088 & G030.0209+0.1048 & \nodata & \nodata & 3092	  & Diffuse HII region & G030.026+00.109 &  K      \\
1156 & 30.0325 &  0.1079 & HIGALBM30.0323+0.1082 & \nodata          & \nodata & \nodata & \nodata & \nodata            & \nodata         & \nodata \\ 		
1157 & 30.0558 & -0.3390 & \nodata				 & \nodata          & \nodata & \nodata & \nodata & \nodata            & G030.055-00.339 &	C      \\
1158 & 30.0670 &  0.0976 & HIGALBM30.0650+0.0991 & \nodata          & \nodata & \nodata & \nodata & \nodata            & G030.067+00.097 &	Q      \\
1159 & 30.0954 &  0.0440 & HIGALBM30.0958+0.0440 & \nodata          & \nodata & \nodata & \nodata & \nodata            & G030.090+00.044 &	Q      \\
1160 & 30.1139 & -0.5646 & \nodata				 & \nodata          & \nodata & \nodata & \nodata & \nodata            & \nodata         & \nodata \\ 		
1161 & 30.1319 & -0.6586 & HIGALBM30.1324-0.6578 & \nodata          & \nodata & \nodata & \nodata & \nodata            & \nodata         & \nodata \\		
1162 & 30.1971 &  0.3085 & HIGALBM30.1971+0.3101 & G030.1963+0.3107 & \nodata & \nodata & \nodata & \nodata            & G030.197+00.309 &	C      \\
\enddata
\tablecomments{This table is available in its entirety in machine-readable format. A portion is shown here for guidance regarding its form and content.}
\end{splitdeluxetable}

\begin{deluxetable}{lccccc}
\tablecaption{YB Position, Velocity, and Bayesian Distance \label{tab:ybvel}}
\tablewidth{0pt}
\tabletypesize{\normalsize}
\tablehead{\colhead{YB} & \colhead{Gal. Longitude}       & \colhead{Gal. Latitude}       & \colhead{Velocity\tablenotemark{a}}    & \colhead{Distance\tablenotemark{b}} & \colhead{$\sigma_{\mathrm{dist}}$} \\
\colhead{Number} & \colhead{(degrees)} & \colhead{(degrees)} & \colhead{(km s$^{-1}$)} & \colhead{(kpc)} & \colhead{kpc}
}
\startdata
 1153 & 30.0033 & -0.2658 & 103.0 &  7.43 & 0.61 \\
 1154 & 30.0233 & -0.0428 &  93.4 &  6.70 & 1.49 \\
 1155 &	30.0242 &  0.1085 &	106.4 &	 7.32 &	0.62 \\
 1156 &	30.0325 &  0.1079 &	106.4 &	 7.31 &	0.62 \\
 1157 &	30.0558 & -0.3390 &	 70.9 &	 4.32 &	0.28 \\
 1158 &	30.0670 &  0.0976 &	 96.6 &	 6.00 &	0.96 \\
 1159 &	30.0954 &  0.0440 &	105.3 &	 7.32 &	0.63 \\
 1160 &	30.1139 & -0.5646 &	101.9 &	 6.36 &	1.53 \\
 1161 &	30.1319 & -0.6586 &	 80.0 &	 4.48 &	0.28 \\
 1162 &	30.1971 &  0.3085 &	  8.0 &	13.50 &	0.35 \\
\enddata
\tablenotetext{a}{YBs with no $^{13}$CO association ($n=23$) indicated by \nodata}
\tablenotetext{b}{Bayesian distance calculator output using P$_\mathrm{far} = 0.5$}
\tablecomments{This table is available in its entirety in machine-readable format. A portion is shown here for guidance regarding its form and content.}
\end{deluxetable}

\begin{splitdeluxetable}{lllccccBccccccBcccccccccc}
\tablecaption{Physical Properties of YB--Hi-GAL Matched Sources \label{tab:ybhigal}}
\tablewidth{0pt}
\tablehead{ \colhead{YB} & \colhead{$l$}  & \colhead{$b$} & \colhead{Distance} & \colhead{$\sigma_{d}$} & \colhead{Hi-GAL Distance} & \colhead{Hi-GAL Dist. flag} & \colhead{Rescaled Mass} & \colhead{$\sigma_M$} & \colhead{Rescaled L$_\mathrm{BOL}$} & \colhead{$\sigma_L$} & \colhead{Rescaled Diameter} & \colhead{$\sigma_D$}  & \colhead{L$_\mathrm{bol}$/Mass} & \colhead{$\sigma_{L/M}$} &  \colhead{T$_\mathrm{grey}^\tablenotemark{a}$} & \colhead{$\sigma_{T_g}$} &  \colhead{L$_\mathrm{ratio}^\tablenotemark{b}$} & \colhead{$\sigma_{L_r}$} & \colhead{T$_\mathrm{bol}^\tablenotemark{c}$} & \colhead{$\sigma_{T_b}$} &\colhead{Surface Density ($\Sigma)$} & \colhead{$\sigma_\Sigma$} \\ 
\colhead{Number} & \colhead{(degrees)} & \colhead{(degrees)} & \colhead{(kpc)} & \colhead{(kpc)} &  \colhead{(kpc)} & \colhead{(G/B/N)} & \colhead{(M$_\sun$)} & \colhead{(M$_\sun$)} &  \colhead{(L$_\sun$)} & \colhead{(L$_\sun$)} & \colhead{(pc)} & \colhead{(pc)} & \colhead{(L$_\sun$/M$_\sun$)} & \colhead{(L$_\sun$/M$_\sun$)} & \colhead{(K)} & \colhead{(K)} & \colhead{} & \colhead{} & \colhead{(K)} &  \colhead{(K)} & \colhead{(g cm$^{-2}$)} & \colhead{(g cm$^{-2}$)}
} 
\startdata
1153 & 30.0033 & -0.2659 & 7.43 & 0.61 & 8.187 & B & 1726.85 & 327.18 & 8055.78 & 1322.75 & 0.49 & 0.04 & 4.67 & 1.17 & 15.59 & 0.24 & 59.87 & 11.97 & 45.59 & 9.12 & 1.886 & 0.36 \\
1154 & 30.0233 & -0.0429 & 6.70 & 1.49 & 8.976 & B & 110.71 & 120.19 & 11118.11 & 5422.01 & 0.51 & 0.11 & 100.42 & 119.52 & 36.44 & 3.56 & 236.91 & 47.38 & 42.53 & 8.51 & 0.112 & 0.12 \\
1155 & 30.0243 & 0.1085 & 7.32 & 0.62 & 7.840 & B & 916.74 & 202.75 & 12669.02 & 2146.12 & 0.65 & 0.06 & 13.82 & 3.85 & 17.08 & 0.66 & 139.1 & 27.82 & 60.22 & 12.04 & 0.584 & 0.13 \\
1156 & 30.0325 & 0.1079 & 7.31 & 0.62 & 7.829 & B & 1010.23 & 326.74 & 6279.24 & 1065.15 & 0.66 & 0.06 & 6.22 & 2.27 & 16.18 & 0.96 & 70.86 & 14.17 & 53.33 & 10.67 & 0.616 & 0.2 \\
1158 & 30.0670 & 0.0976 & 6.00 & 0.96 & 1.000 & N & 261.36 & 85.26 & 2995.92 & 958.69 & 0.40 & 0.06 & 11.46 & 5.24 & 18.62 & 0.34 & 95.11 & 19.02 & 44.64 & 8.93 & 0.443 & 0.14 \\
1159 & 30.0955 & 0.0441 & 7.32 & 0.63 & 7.923 & B & 714.58 & 174.29 & 2504.42 & 431.09 & 0.63 & 0.05 & 3.5 & 1.05 & 14.34 & 0.48 & 53.44 & 10.69 & 49.33 & 9.87 & 0.485 & 0.12 \\
1161 & 30.1320 & -0.6587 & 4.48 & 0.28 & 4.963 & G & 116.99 & 114.08 & 461.02 & 108.73 & 0.61 & 0.04 & 3.94 & 3.95 & 16.00 & 0.72 & 45.65 & 9.13 & 40.26 & 8.05 & 0.083 & 0.08 \\
1162 & 30.1971 & 0.3085 & 13.5 & 0.35 & 14.156 & B & 2034.84 & 114.10 & 20209.16 & 1047.88 & 0.71 & 0.02 & 9.93 & 0.76 & 19.22 & 0.19 & 78.23 & 15.65 & 47.19 & 9.44 & 1.083 & 0.06 \\
\enddata
\tablenotetext{a}{Dust temperature calculated using greybody fit}
\tablenotetext{b}{Ratio of L$_\mathrm{bol}$ to luminosity calculated for $\lambda \geq 350~\mu$m. }
\tablenotetext{c}{Bolometric temperature calculated using equation (5) in EMS17.}
\tablecomments{This table is available in its entirety in machine-readable format. A portion is shown here for guidance regarding its form and content.}
\end{splitdeluxetable}

\begin{splitdeluxetable}{lcccB lccccc}
\tablecaption{Descriptive Statistics for the YB--Hi-GAL Matched Sources \label{tab:ybhigalstats}}
\tablewidth{0pt}
\tablehead{\colhead{} & \colhead{Rescaled Mass}             & \colhead{Rescaled Luminosity}       & \colhead{Rescaled Diameter} & \colhead{} & \colhead{ L$_\mathrm{bol}$/Mass} & \colhead{T$_\mathrm{grey}$} & \colhead{L$_\mathrm{ratio}$} & \colhead{T$_\mathrm{bol}$} & \colhead{Surface Density} \\
           \colhead{} & \colhead{$\log\left(M_\sun\right)$} & \colhead{$\log\left(L_\sun\right)$} & \colhead{$\log\left(\mathrm{pc}\right)$} & \colhead{} & \colhead{$\log\left(M_\sun/L_\sun \right)$} & \colhead{(K)} & \colhead{$\log\left(L/L_\mathrm{smm}\right)$} & \colhead{(K)} & \colhead{$\log\left(\mathrm{g~cm}^{-2}\right)$}
}
\startdata
Median  & 2.51 & 3.33 & $-0.32$ & Median &  0.96  &  17.8 &  1.91 & 47.0  & $-0.49$ \\
Mean    & 2.37 & 3.30 & $-0.37$ & Mean   &  0.93  &  18.8 &  1.89 & 47.2  & $-0.46$ \\
SD      & 0.80 & 0.93 & 0.33 & SD     &  0.58  &  5.1  &  0.37 & 9.8   & 0.52  \\
Max.    & 4.20 & 6.05 & 0.33 & Max.   &  2.36  &  40.0 & 2.74  & 127.4 & 0.83  \\
Min.    & 0.08 & 0.53 & $-1.30$ & Min.   & $-1.0$ &  8.9  & 0.79  & 18.4  & $-2.3$  \\
\enddata

\end{splitdeluxetable}

\subsection{Clump Physical Properties} \label{sec:props}

In this section we examine the properties of the YBs that have associations with EMS17 Hi-GAL compact sources/clumps (see \autoref{sec:cats} for details on the catalog cross-match procedure). These sources are of particular interest as they are the subset of YBs most likely to be associated with star-formation activity across a range of both intermediate and high masses. We omit 17 of the 385 YB--Hi-GAL matched sources as they do not have a $^{13}$CO association and related distance measurement. Data for the remaining 368 YB--Hi-GAL sources are shown in \autoref{tab:ybhigal}. Columns 1 through 3 give the YB identification and positional information, and columns 4 through  7 give distance-related data as discussed in \autoref{sec:distances}.  Columns 8 through 13, give the EMS17 mass, bolometric luminosity, and diameter, rescaled using the distance in column 4. The remaining  columns (14--23) give distance-independent quantities calculated by EMS17.

We calculated the mass uncertainty (column 9) by combining the uncertainty in the mass (DMASS) listed in EMS17, and our distance uncertainty in column 5. The Hi-GAL masses were derived as SED fit parameters, and the uncertainty (DMASS) is the larger of $|M-M_\mathrm{low}|$ and $|M-M_\mathrm{high}|$, where $M$ is the best-fit mass, and  $M_\mathrm{low}$ and $M_\mathrm{high}$ are the minimum and maximum mass estimates associated with an acceptable SED fit (D. Elia, 2020, personal communication). Two different SED fitting techniques (denoted ‘thick’ and ‘thin’) were used by EMS17. In the majority of the thin fits the clump mass was not highly constrained by the SED fit (e.g., YB 1154 and YB 1161); in such cases the masses are essentially known to within a factor of two.

The bolometric luminosity uncertainty (column 11) arises from uncertainties in the SED integration and the source distance. Roughly the \emph{maximum} uncertainty in the SED integration is thought to be $\sim 20$\% (D. Elia, 2020, personal communication). For sources with a poor SED, which we defined as those having a relative mass uncertainty $>0.75$, we combined a 20\% error with our distance uncertainty. For other sources we only report the uncertainty associated with our distance estimate.

The remaining uncertainties listed in \autoref{tab:ybhigal} are more straightforward to explain. The rescaled diameter uncertainty in column 13 reflects our distance uncertainty. The L/M uncertainty was calculated by propagating the rescaled mass and rescaled bolometric luminosity uncertainties from columns 9 and 11. The $T_\mathrm{grey}$ uncertainty in column 17 was listed in EMS17 and is repeated here. As both the $L_{\mathrm{ratio}}$ and $T_\mathrm{bol}$ quantities involve the integration of SEDs, we report a 20\% uncertainty as a conservative error estimate. Finally, the surface density error reflects the relative mass uncertainty.

\begin{figure}
    \centering
    \includegraphics{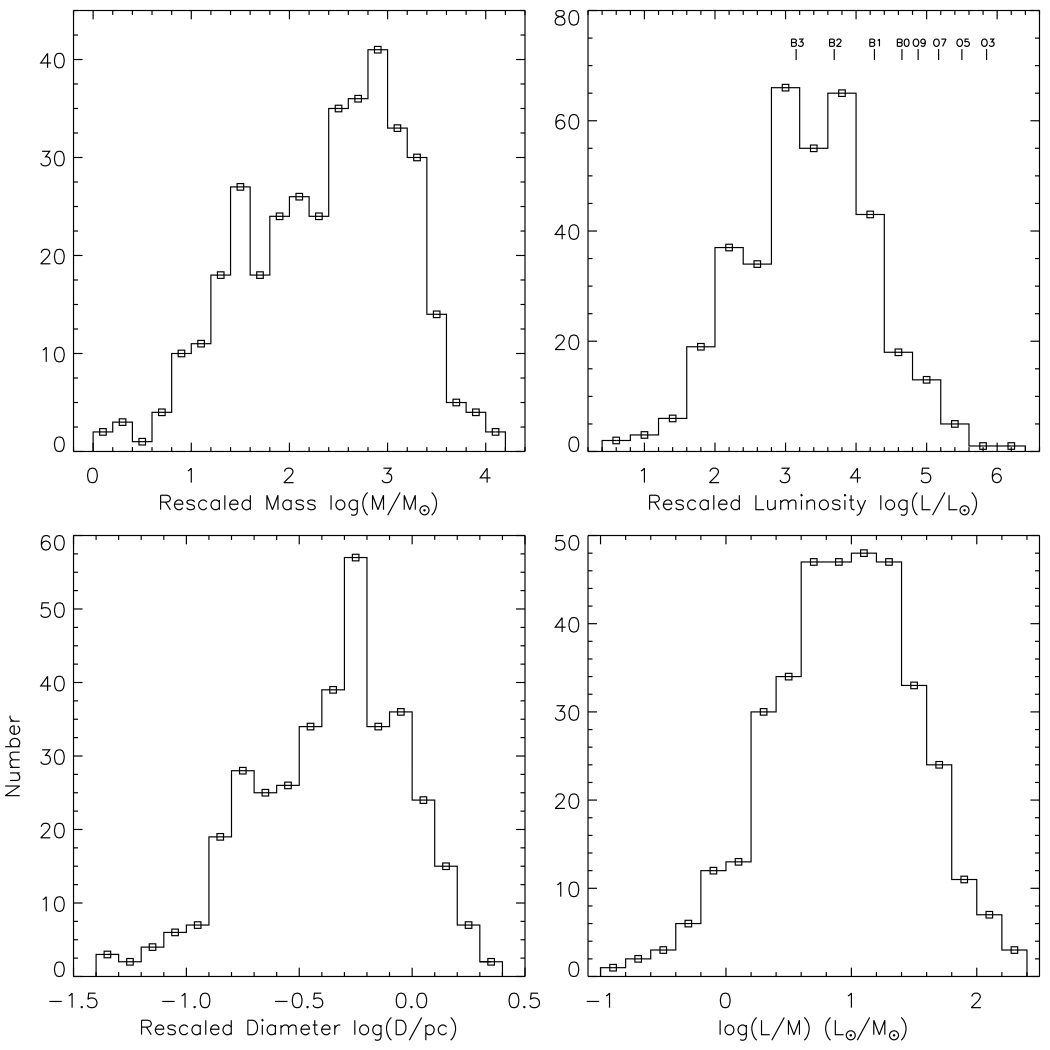}
    \caption{Physical properties of the YB--Hi-GAL matched sources. The lower-left and upper row histograms show physical properties from the EMS17 catalog rescaled using our newly calculated distances. For reference, main-sequence OB star luminosities \citep{cro05} are indicated in the upper-right panel. The lower-right histogram displays the distance-independent luminosity-mass ratio also from EMS17. See text for full discussion.}
    \label{fig:higalhisto1}
\end{figure}

\begin{figure}
    \centering
    \includegraphics{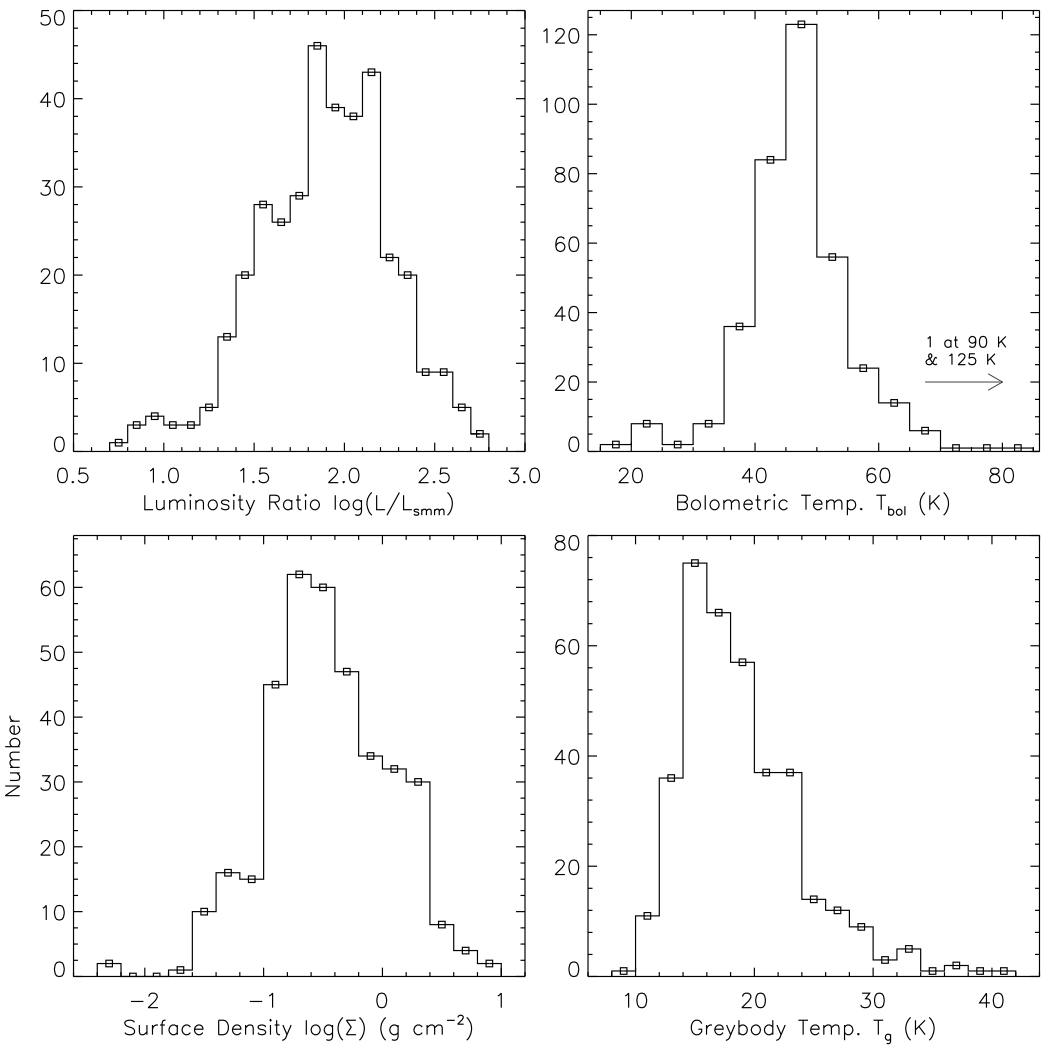}
    \caption{As Figure~\ref{fig:higalhisto1}. All quantities plotted are distance independent.}
    \label{fig:higalhisto2}
\end{figure}

Descriptive statistics for the clump properties are given in \autoref{tab:ybhigalstats}, and histograms of the quantities are shown in \autoref{fig:higalhisto1} and \autoref{fig:higalhisto2}. We note that the diameter reported in \autoref{tab:ybhigal} more closely represents the size of the star-forming region containing the YB and may overestimate the size of the YB itself. We address the issue of YB size in more detail in \autoref{sec:fits} and \autoref{sec:size}.

EMS17 divided the Hi-GAL compact sources into two major groups, prestellar and protostellar. The prestellar sources showed no signs of star-formation activity, while protostellar sources had emission at 70~$\mu$m, indicative of embedded star formation activity. As expected, 97\% of YB-matched HiGal sources were flagged by EMS17 as protostellar, and we further assume that the remaining Hi-Gal matched YBs are also protostellar because they are emitting in the MIR. Comparison with figures from EMS17 show that the mass, luminosity, and diameter distributions shown in \autoref{fig:higalhisto1} are all consistent with the properties expected for protostellar objects. This is not unexpected as YBs, by definition, have MIR emission that is likely due to embedded star-formation activity. Referring to \autoref{fig:higalhisto1}, we see the YB--Hi-GAL source masses ($M$) fall mostly between $0 < \log(M) < 4$, and the source luminosity ($L$) falls between $ 1 < \log(L) < 5$.  Most of the YB--Hi-GAL sources have diameters ($D$)  that are in the canonical ``clump" range ($0.2 \leq D \leq 3$ pc), and thus likely contain multiple sites of star formation. Our diameter distribution is shifted slightly to lower diameters compared to the overall Hi-GAL sample (cf. Figure 4 in EMS17), and thus includes a larger proportion of ``cores" ($D < 0.2$ pc or $\log(D) < -0.70$). This slight difference may be because of differences in how we resolved the near/far distance ambiguity for sources compared to EMS17. As discussed in \autoref{sec:distances}, we have assigned a near distance for several sources in our pilot region that EMS17 assumed to be at a far distance. 

The distance-independent quantities provide additional information about the the evolutionary stage of the embedded star-formation activity as well as the mass of the forming stellar population. Comparing the lower-right panel of \autoref{fig:higalhisto1} with Figure 13 in EMS17, we see that the range of the bolometric luminosity to mass ratio ($L/M$) is consistent with protostellar sources, but  the YB--Hi-GAL peak is shifted to a larger value ($\log(L/M) = 0.96$ cf. $\log(L/M) \sim 0.4$ ). EMS17 use $L/M \geq 22.4$ $\mathrm{L}_\sun/\mathrm{M}_\sun$ (or $\log(L/M) \geq 1.35$) to define ``\ion{H}{2}-region candidates", i.e., compact sources that have embedded high-mass star formation activity. This makes up about 10\% of their protostellar sample. In contrast, 24\% of the YB--Hi-GAL sources have ($L/M$) consistent with being \ion{H}{2}-region candidates, which is indicative of YBs being associated with a mix of intermediate- and high-mass star formation. 

The ratio of the bolometric luminosity to the luminosity in the sub-millimeter ($\lambda \geq 350~\mu$m) can be used as a rough proxy for the evolutionary stage of the star-formation activity. Regardless of the details of the star-formation process, as the source evolves, and more stars form within the clump/core, an increasing proportion of the luminosity will originate at shorter wavelengths. EMS17 chose $\log(L/L_\mathrm{smm}) < 2$ as a representative dividing line between early and more evolved star formation, mirroring the criteria used to define low-mass, isolated, Class 0 young stellar objects. Comparison of the upper left panel in \autoref{fig:higalhisto2} with Figure 14 in EMS17 shows that while the range of $\log(L/L_\mathrm{smm})$ is similar to the protostellar sample, the YB--Hi-GAL sample is shifted to the right with 40\% of the YB--Hi-GAL sample having $\log(L/L_\mathrm{smm}) \geq 2$ compared to only 14\% for the full protostellar sample. This shows YB--Hi-GAL sample is slightly more evolved than the full Hi-GAL sample. 

The bolometric temperature (T$_\mathrm{bol}$), which is defined as the temperature of a blackbody that has the same mean frequency as the observed spectral energy distribution (SED; \citealt{myers93}), is another quantity that can easily be related to the evolutionary stage of the object. A cooler T$_\mathrm{bol}$ would correspond to earlier stages of star formation where the peak of the SED is shifted to longer wavelengths. Comparing the upper right panel of \autoref{fig:higalhisto2} with Figure 15 in EMS17, we see that the YB--Hi-GAL sources clearly fall in the protostellar range, but there is a deficit of YB--Hi-GAL sources with T$_\mathrm{bol} <40$~K (only 15\% compared to $\sim50$\% for the full sample). This is consistent with the results from the luminosity ratio outlined in the previous paragraph indicating the YB--Hi-GAL sample is slightly more evolved than the full Hi-GAL sample. 

The surface density of a source is often used to identify regions associated with high-mass star formation. A comparison of the lower left panel of \autoref{fig:higalhisto2} with the protostellar sample shown in Figure 7 of EMS17 shows that both distributions have the same peak, around $\log(\Sigma) = -0.5$, and similar ranges; however, the YB--Hi-GAL distribution is not as symmetric, and it is slightly shifted to higher surface densities. This shift can be quantified by looking at the fraction of sources found above the \citet{km08} threshold for high-mass star formation;  21\% of the YB-Hi-GAL sample have $\Sigma > 1 $ g~cm$^{-2}$ compared with 13\% for the full protostellar sample.

Finally, the temperature used in the greybody fits to the Herschel SEDs at $\lambda \geq 160 \mu$m (T$_\mathrm{grey}$) is shown in the lower right panel of \autoref{fig:higalhisto2}. The distribution is similar to the protostellar T$_\mathrm{grey}$ distribution shown in Figure 5 of EMS17. The YB--Hi-GAL distribution has a slightly higher average temperature (18.8~K)  than the protostellar sample (16.0~K), but it shares approximately the same low temperature cutoff and high temperature tail.

The comparisons outlined in this section clearly show that the YB--Hi-GAL sources are protostellar objects associated with intermediate- and high-mass star formation spanning a range of evolutionary stages, and are more massive and slightly more evolved than the full Hi-GAL sample. In \autoref{sec:samps} we combine various observables to learn more about the YB--Hi-GAL sample, and we look at the physical properties of subsets of this sample. 

\subsection{MIR Photometry} \label{sec:photo}

We obtained MIR photometry of YBs at 8.0, 12.0, and 24~$\mu$m from our \emph{Spitzer} and \emph{WISE} images for 511 of the 516 sources in our pilot region, with five sources omitted due to incomplete image coverage of the source. The Galactic plane exhibits complex, structured backgrounds at MIR wavelengths. Additionally, YBs are typically extended sources, and many YBs that are single, circular features at 12 and 24 $\mu$m contain structure like filaments or multiple sources at 8 $\mu$m. This is not surprising, due to the higher resolution at 8 $\mu$m, and the expectation that young PDRs contain multiple seeds of star formation. Thus both traditional ``stellar'' aperture photometry, and PSF-fitting photometry are not appropriate techniques. Instead, photometry was done using a program written in Python inspired by the IDL-based {\sc imview} program \citep{hig97}. Our  program allows the user to display an image and interactively select points surrounding the YB that are used to define the local background. The program then interpolates over the area within those points using a multiquadratic radial basis function, creating a background estimate. The difference between background and image is calculated, resulting in a residual ``YB-only" image, and converted to a flux density. Photometry was not conducted for sources that were saturated at 12.0 and/or 24~$\mu$m.

The largest source of uncertainty in our flux-density measurements comes from the choice of points used to define the background. To address this uncertainty three different astronomers conducted the photometry twice for a total of six photometric measurements of each source. We report the mean and fractional error (standard deviation divided by mean) of these measurements in \autoref{tab:ybphot}. We only included sources with fractional error of $< 50\% $ in our later analysis in \autoref{sec:colors}. The Hi-GAL 70~$\mu$m photometry results of EMS17 are also included in \autoref{tab:ybphot} for convenience. EMS17 include additional photometry at longer wavelengths, but these longer wavelengths become increasingly less likely to accurately be associated with the YBs so we do not include them here. 

While the basic technique was the same for all three bands, there were some instrument-specific considerations. IRAC images were in units of MJy sr$^{-1}$, so obtaining the flux density for each source required only a pixel-to-steradian conversion factor of $3.385\times10^{-11}$ steradians pixel$^{-1}$ (1.2 arcsec pixels)  \citep{faz04}. The IRAC Instrument Handbook \citep{IRAC2015} includes plots showing recommended aperture corrections for 8.0~$\mu$m photometry of extended sources. The average MWP radius for the DR2 YBs is $\sim12$ arcsec, and, if we take this to be a typical aperture size, an aperture correction of at most 0.85 -- 0.8 should be applied. To simplify the collection of photometry we chose not to apply any aperture corrections to our data as the resulting shift in the MIR colors is very small. For example, the largest aperture correction expected ($\sim 0.8$) would correspond to a shift in $F_{12}/F_{8}$ of only 0.1 dex. A source radius may be used to select an aperture correction for the data shown in \autoref{tab:ybphot}, however this would introduce new uncertainty from ambiguity in appropriate source radius (see discussion in \autoref{sec:size}). The \emph{WISE} Allsky Catalog release document \citep{cut12} lists a \emph{WISE} 12-$\mu$m band aperture correction for extended objects of 0.973. As with the IRAC data, we did not apply the correction. The raw photometry in WISE data numbers (DN) was converted to Jy using a DN-to-Jy conversion factor of $1.8326\times10^{-6}$ Jy DN$^{-1}$, which incorporates recommended zero point magnitudes and fluxes. Finally, the MIPS instrument handbook \citep{MIPS2011} did not suggest any aperture corrections for extended sources. MIPS images were in units of MJy sr$^{-1}$ so conversion to flux density involved only the pixel-to-steradian conversion factor of $3.673\times10^{-11}$ steradians pixel$^{-1}$ (1.25 arcsec pixels).

\begin{deluxetable}{r cc cc cc cc c} 
\tablecaption{Pilot Region MIR Photometry \label{tab:ybphot}} 
\tablewidth{0pt}
\tabletypesize{\normalsize} 
\tablehead{\colhead{Source} & \colhead{$F_8$\tablenotemark{a}} & \colhead{$\frac{\sigma}{F_8}$\tablenotemark{a}}  & \colhead{$F_{12}$\tablenotemark{a}} & \colhead{$\frac{\sigma}{F_{12}}$\tablenotemark{a}}  & \colhead{$F_{24}$\tablenotemark{a}}  & \colhead{$\frac{\sigma}{F_{24}}$\tablenotemark{a}} & \colhead{$F_{70}$\tablenotemark{b}}  & \colhead{$D F_{70}$\tablenotemark{b}} & \colhead{Well-fit by 2D Gaussian Model} \\  
 & \colhead{(Jy)} &  & \colhead{(Jy)} & & \colhead{(Jy)} & & \colhead{(Jy)} & \colhead{(Jy)}   &  }
  \startdata 
1153  &  0.28  &  0.23  &  0.05  &  0.29  &  0.15  &  0.13  &  64.85  &  4.1 & x \\  
1154  &  15.68  &  0.22  &  7.68  &  0.29  & -- & -- &  327.27  &  19.97 & \\  
1155  &  1.36  &  0.29  &  0.67  &  0.13  &  2.96  &  0.11  &  64.34  &  6.5 &  \\  
1156  &  0.75  &  0.15  &  0.16  &  0.25  &  1.09  &  0.18  &  46.17  &  1.72 & \\  
\enddata 
\tablenotetext{a}{We report the mean and fractional error (standard deviation divided by mean) of six separate photometric measurements of each source. See text for details.}
\tablenotetext{b}{Values reported by EMS17.}
\tablecomments{This table is available in its entirety in machine-readable format. A portion is shown here for guidance regarding its form and content.} 
\end{deluxetable}

\subsection{2D-Gaussian Fits} \label{sec:fits}

The YBs in the full catalog generated by MWP DR2 are very heterogeneous, with many showing extended structure and multiple point sources. To isolate compact structures, we fit 2D Gaussians to the YB 8-$\mu$m and 24-$\mu$m emission in the background-subtracted, ``YB-only" images generated in the photometry process discussed in \autoref{sec:photo}.  We note that while the photometry was repeated six times per source, only one set of background-subtracted images was used for the fitting. To see how well-fit the YB-only images were by a 2D Gaussian, we created a residual of 2D-Gaussian fit subtracted from the YB-only image, which we refer to as the ``residual image" in the rest of this section. We measured the magnitude of residuals, differences in 8-$\mu$m and 24-$\mu$m peak locations, and differences in 8-$\mu$m and 24-$\mu$m Gaussian standard deviations. Coincidence of 8-$\mu$m and 24-$\mu$m peaks and comparable spatial extents at these wavelengths are expected features of young PDRs. In order to assess the similarity of the fits at 8 $\mu$m and 24 $\mu$m, we defined a series of points. YBs were given up to 4 points, one point for fulfilling each of the following criteria:

\begin{itemize}
\item[] (1) The absolute value of the flux in the residual is $\le$10\% of the flux in the  Gaussian fit to the 8-$\mu$m image.
\item[] (2) The absolute value of the flux in the residual is $\le$10\% of the flux in the  Gaussian fit to the 24-$\mu$m image.
\item[] (3) The distance between the center of the 8-$\mu$m and 24-$\mu$m Gaussians is $<$0.3 times the average of the 8-$\mu$m and 24-$\mu$m standard deviations $(\sigma_{8\mu mx}+\sigma_{8\mu my}+\sigma_{24\mu m x}+\sigma_{24\mu my})/4$.
\item[] (4) The normalized difference between the average of the 8-$\mu$m and 24-$\mu$m standard deviations satisfies $|\sigma_{8\mu m} - \sigma_{24\mu m}|$/(($\sigma_{8\mu m} + \sigma_{24\mu m})/2) < 0.4$.

\end{itemize}

Each source that gained four points can therefore be considered well-fit by a 2D-Gaussian model at both  8-$\mu$m and 24-$\mu$m. These Gaussian sources are consistent with being compact PDRs or possibly evolved stars; however, our color analyses suggest that YBs are predominantly young sources (see \autoref{sec:colors}).  Ninety-six of the full sample of 516 YBs received four points (18.6\%): 65 of these have Hi-GAL clump matches; 9 have only WISE (8Q, 1C) catalog matches; 6 have no associated molecular gas velocities; and 16 lack matches in any catalog. Sources that earned four points are flagged in \autoref{tab:ybphot}. In \autoref{sec:size} we compare the results of our Gaussian fits with MWP YB sizes and the sizes of associated Hi-GAL clumps. 

\section{Analysis \& Discussion} \label{sec:analysis}

\subsection{Properties of YB--Hi-GAL Subsets} \label{sec:samps}

\begin{figure}
    \centering
    \includegraphics{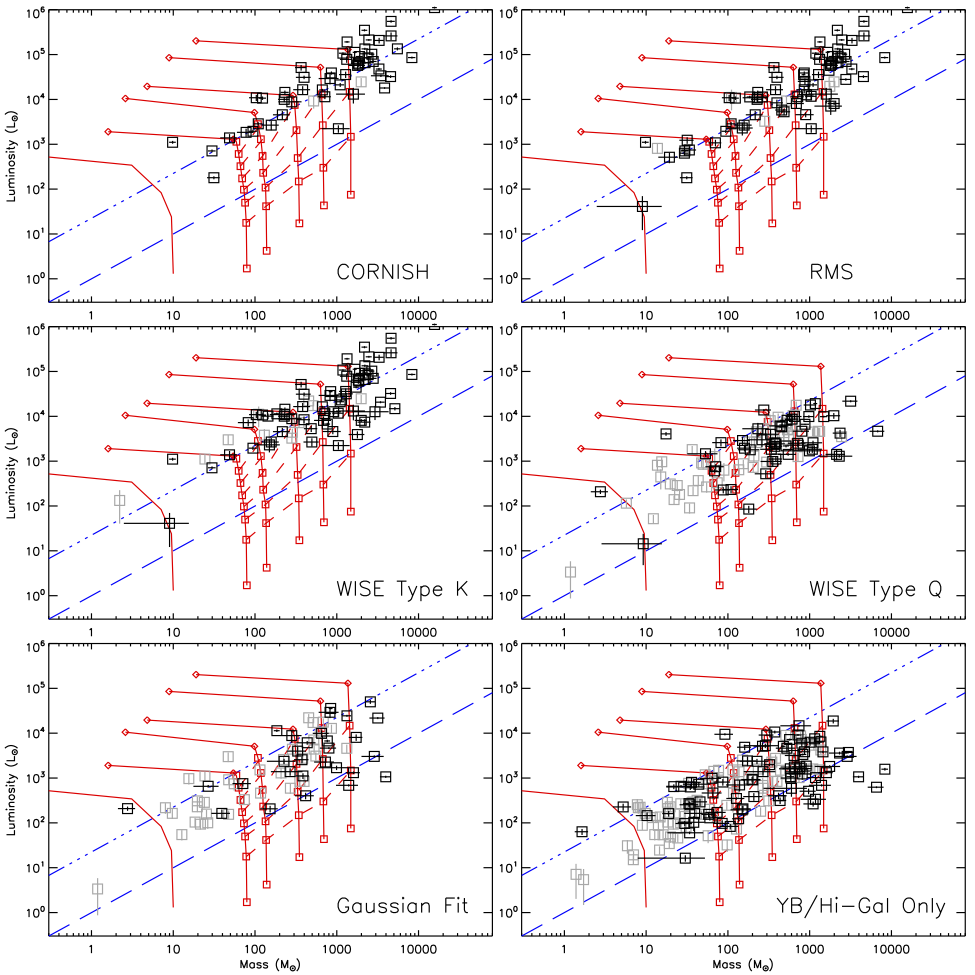}
    \caption{Bolometric Luminosity -- Clump Mass plots for various subsets of the YB--Hi-Gal Sources. Red lines show \citet{mol08} clump evolutionary tracks. Small squares, along the vertical portion of each track, correspond to $5 \times10^4$ year intervals, and isochrones are the red short dashed lines. The blue upper diagonal line represents a 90$^{\mathrm{th}}$ percentile lower limit for \emph{Herschel}-defined \ion{H}{2} regions, and the blue lower diagonal line is a 90$^{\mathrm{th}}$ percentile upper limit for the pre-stellar objects in the EMS17 sample of Hi-GAL clumps(see text for details). For sources with good SED fits (black), mass and luminosity error bars are shown (most are smaller than the symbol used). Sources with poor SED fits (grey) do not have error bars shown for mass.}
    \label{fig:blplots}
\end{figure}

In \autoref{fig:blplots} we show bolometric luminosity (L) vs. clump, or envelope, mass (M) plots for different subsets of YB--Hi-GAL matched clumps. The position of a clump in an LM-plot is related to its evolutionary stage. On each plot we show simplified evolutionary tracks based on the \citet{mol08} model for the evolution of star-forming protostellar clumps, and each track corresponds to a star-forming clump with given initial envelope mass. The vertical portions of each track correspond to when the embedded protostar(s) are forming and are rapidly gaining mass from their birth clouds; the luminosity of the cloud increases due to accretion and internal heating from the protostar(s), and the gas mass of the clump is lowered slightly reflecting primarily the conversion of gas to stars. The subsequent horizontal tracks (between diamond symbols) correspond to the relatively slower process of cloud dispersal; the luminosity remains roughly constant, reflecting the luminosity of the newly formed stars, while the remaining gas is removed via processes like bipolar outflows, stellar winds, radiation pressure, and photoionization. Small squares along the vertical portion of  each line correspond to $5 \times10^4$ year intervals, and isochrones are indicated using a short dashed line.  The upper diagonal line represents a 90$^{\mathrm{th}}$ percentile lower limit for \ion{H}{2} regions, $\left(L/M\right)_\mathrm{HII} \geq 22.4$ L$_\sun$/M$_\sun$, which was defined by EMS17 using Hi-GAL--CORNISH \ion{H}{2} region matches. The lower diagonal line is a 90$^{\mathrm{th}}$ percentile upper limit for the pre-stellar objects in the EMS17 sample of Hi-GAL clumps,  $\left(L/M\right)_\mathrm{Pre} \leq 0.9$ L$_\sun$/M$_\sun$.

Not surprisingly, we see that the vast majority of the CORNISH-, RMS-, and  known compact \ion{H}{2} regions (WISE-K)-matched YBs fall primarily along and above the $\left(L/M\right)_\mathrm{HII}$ line. They also tend to be near the vertical-horizontal transition region for the various evolutionary tracks as would be expected for objects like compact \ion{H}{2} regions, which represent fairly evolved states of massive star evolution. Note that there is significant overlap between these three groups (see \autoref{tab:xmatch}). In contrast, objects that are radio-quiet (WISE-Q), well-fit by 2D Gaussians, and YBs with only Hi-GAL matches, tend to lie along the active accreting portion of the evolutionary tracks for a wide range of clump masses and luminosity. There is also a noticeable deficit of these sources (compared to the CORNISH, RMS and WISE-K samples) in the upper-right corner of the LM plots ($M>10^3$~M$_\sun$, $L>10^4$~L$_\sun$).

 WISE-Q sources are particularly interesting as they are objects that were classified as potential \ion{H}{2} regions based on their infrared colors and morphology, but they lack significant/detectable radio emission \citep{and14}. Our examination of the LM plots suggests that a (modest) majority of the WISE-Q objects are radio quiet because they are associated with low- and intermediate-mass star formation. We note that sources associated with clump masses below 500 M$_\sun$ represent about 63\% of the sample. If these objects are evolving along the evolutionary tracks shown, they correspond to regions with a final luminosity of $<10^4$ L$_\sun$, which, assuming that a single star forms, is the luminosity expected for a single B1~V star. The remaining WISE-Q sources ($\sim 37$\%) are associated with more massive clumps that are still on the vertical portion of their evolutionary tracks. In this case the lack of radio emission could be because ionizing photons are being absorbed by the dusty envelope surrounding a newly-formed massive star, or the clump is forming multiple intermediate-mass stars that are unable to form a significant \ion{H}{2} region. We note that this proportion of WISE-Q sources associated with high-mass star formation (37\%) is much higher than the number given using a $\Sigma>1$ g cm$^{-2}$ cutoff (5.3\%); however, it is comparable to the number found using a  $\Sigma>0.3$ g cm$^{-2}$ cutoff (41\%) suggested by \citet{lop2010}. For comparison, the fraction of the entire YB--Hi-GAL sample associated with high-mass star formation is 21\%, 53\%, and 41\% using $\Sigma>1$ g cm$^{-2}$, $\Sigma>0.3$ g cm$^{-2}$, and clump mass $>500$ M$_\sun$ respectively.
 
 The 2D-Gaussian well-fit sources, shown in the lower-left panel of \autoref{fig:blplots}, have a uniform distribution in both clump mass and luminosity. The majority of the sources appear to be in the accretion stage of their evolution. This may be due to the fact that more evolved sources have a more complex infrared morphology formed as material is dispersed.

 The YB--Hi-GAL only sources also have a uniform distribution of clump mass and luminosity, and again the majority of the sources appear to be in the accretion stage of their evolution. Unlike the other subsets, the YB--Hi-GAL only sample includes a very large fraction of sources that are associated with lower-mass clumps (45\% of sources associated with clumps with $M<10^2$~M$_\sun$).
 
\begin{figure}
    \centering
    \includegraphics[width=\textwidth]{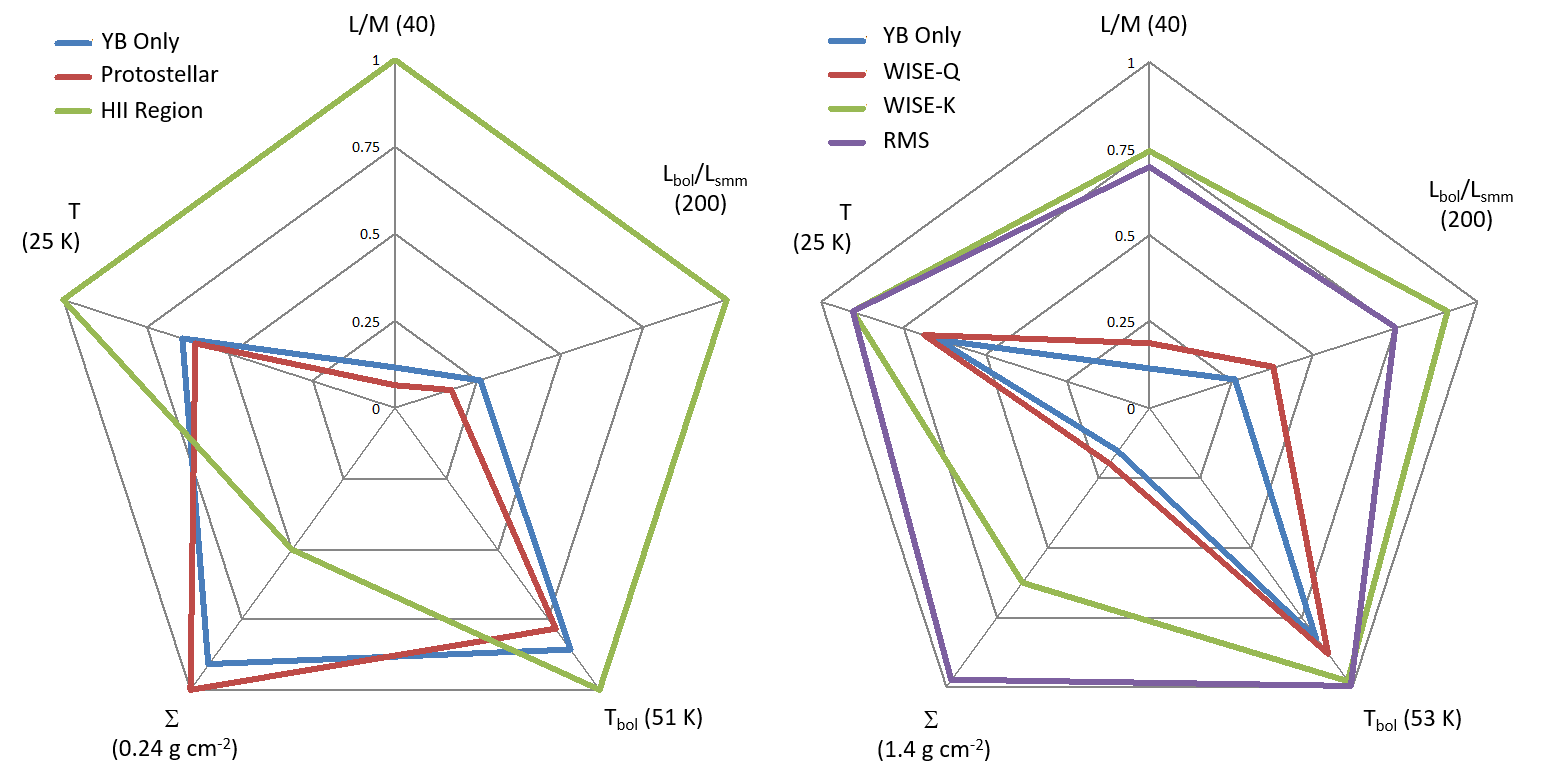}
    \caption{Radar-plot visualizations for five distance-independent quantities for different classes of objects. Values plotted are medians, and the maxima are indicated for each radial line. Different classes are connected via different colored pentagons.  Left: Protostellar clumps and \ion{H}{2} regions identified in the Hi-GAL survey (EMS17) are compared with YB--Hi-Gal matches lacking counterparts in other surveys. Right: Four YB--Hi-GAL subsets are compared. The RMS and WISE-K subsets are almost entirely high-mass star-forming regions, while the WISE-Q and YB--Hi-GAL only subsets are a mix of intermediate-mass regions and young high-mass regions.}
    \label{fig:radar}
\end{figure}

As discussed in \autoref{sec:props}, distance-independent quantities are useful for distinguishing between different evolutionary stages and different masses in star-forming regions. The five quantities we examine are dust temperature (T), bolometric temperature (T$_{bol}$), ratio of bolometric luminosity to luminosity calculated over wavelengths longward of 350 $\mu$m (L$_{bol}$/L$_{submm}$), ratio of bolometric luminosity to clump mass (L$_{bol}$/M), and surface density of the clump ($\Sigma$). One way to simultaneously compare these quantities for different samples is  through a `radar plot'  showing the median values (EMS17). 

\autoref{fig:radar} presents two radar plots of the five distance-independent quantities associated with different subsets of the YB--Hi-GAL sample.  In each case the median of each quantity is plotted along one of five axes. Each axis is normalized to the value indicated at the end of the axis in order to account for the different numerical ranges spanned by each quantity. The left-hand panel compares our YB-only subsample of YB--Hi-GAL sources with the protostellar and \ion{H}{2} region subsamples from EMS17. On average the YB--Hi-GAL and protostellar samples are very similar, and they are both quite distinct from the \ion{H}{2} region sample. The right-hand plot compares the median values for the different subsets defined in this study. We notice that the WISE-K and RMS samples are similar except for the much higher surface density associated with the RMS sample. This likely reflects the more embedded nature of the objects selected for the RMS catalog. If we consider the WISE-K and RMS subsets to be essentially all high-mass regions, the position of the WISE-Q and YB-Only subsets reflects having an increasing proportion of intermediate-mass star-forming regions in the sample; a lower L$_{bol}$/M and $\Sigma$ would be expected. In addition, these subsets could contain a population of less-evolved, higher-mass regions, which would have a lower L$_{bol}$/L$_{submm}$ and a lower T$_{bol}$. 

\subsection{YB Colors} \label{sec:colors}

\begin{figure}
    \centering
    \includegraphics{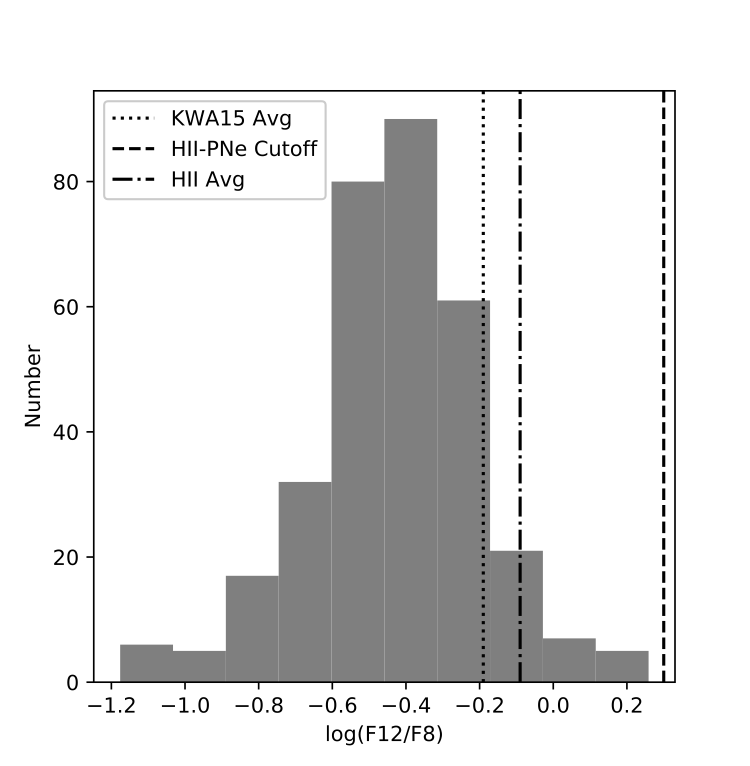}
    \caption{Distribution of the $\frac{F12}{F8}$ colors for the 324 YBs in our pilot region with reliable 8 and 12$~\mu m$ photometry. The vertical dotted line is the average color from KWA15; the dot-dashed line is the average color of \ion{H}{2} regions from \citet{and12}, the dashed line is the color separation \ion{H}{2} region-PNe cutoff from \citet{and12}.}
    \label{fig:colorhist}
\end{figure}

\begin{figure}
    \centering
    \includegraphics{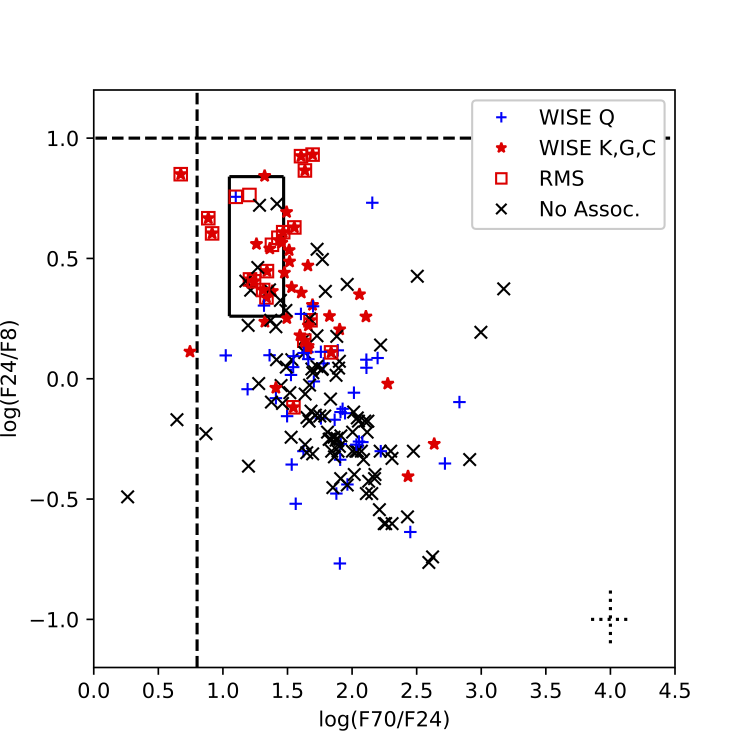}
    \caption{Color-Color plot of the 219 YBs in our pilot region with reliable 8 and 24$~\mu m$ photometry and Hi-Gal counterparts. The 8 and 24$~\mu m$ photometry is described in \autoref{sec:photo}. The 70$~\mu m$ photometry is from EMS17. The colors and shapes indicate YB association with objects in other catalogs, using cross-matching as described in \autoref{sec:cats}. The solid square indicates the region of mean colors of \ion{H}{2} regions established by \citet{and12}, while vertical and horizontal dashed lines indicate the \citet{and12} IR color cut-offs for distinguishing \ion{H}{2} regions from PNe. Dotted lines in lower right indicate the average uncertainty in the reported values.}
    \label{fig:ccp}
\end{figure}

\begin{figure}
    \centering
    \includegraphics[scale=0.6]{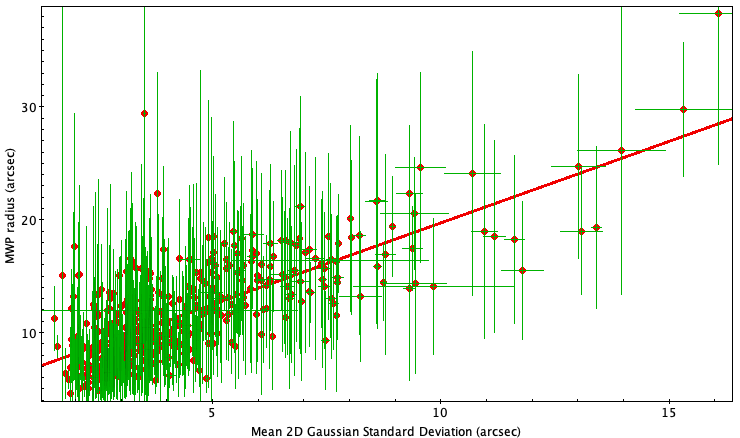}
    \includegraphics[scale=0.6]{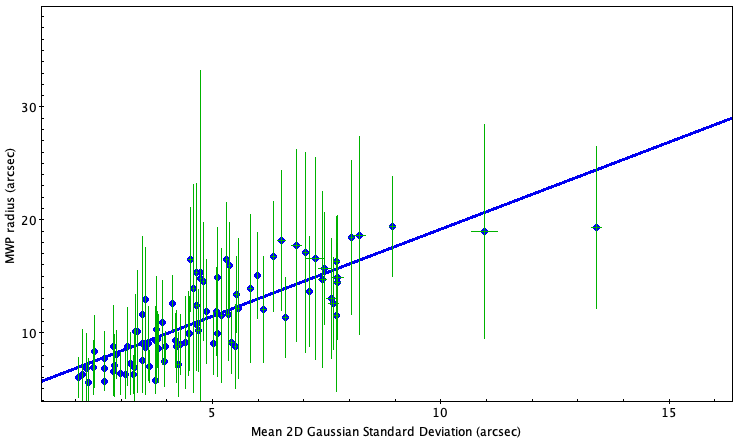}
    \caption{MWP Radius vs. Mean 2D Gaussian Standard Deviation for unsaturated YBs. Top panel: 450 YBs (red symbols and linear fit). Bottom panel: Subset of 89 well-fit YBs (blue symbols and linear fits). Error bars are shown in green. Many of the errors on the Gaussian standard deviations are too small to be visible.}
    \label{fig:YBsizes}
\end{figure}

\begin{figure}
    \centering
    \includegraphics[width=\textwidth]{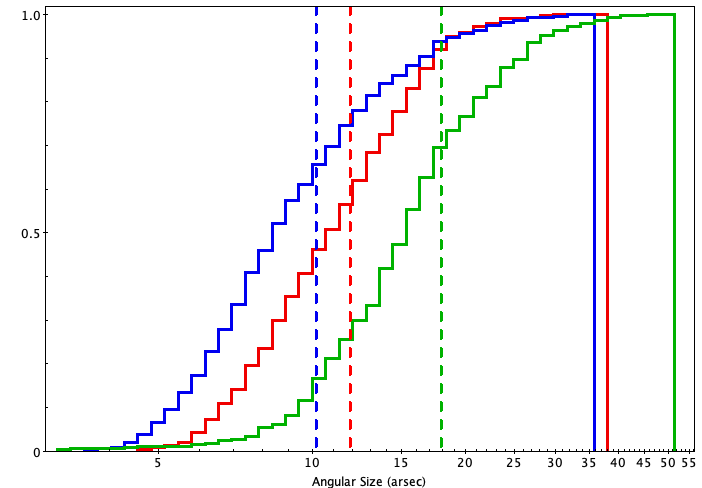}
    \caption{ Normalized cumulative histograms for user-measured MWP radii (red), 2D-Gaussian FWHM (blue), and Hi-GAL 250-$\micron$ FWHM (green) YB size distributions. Dashed vertical lines indicate the mean angular sizes for each distribution, using the same colors.}
    \label{fig:sizecumul}
\end{figure}

A significant finding of KWA15 was that IR colors of YBs, as measured by their fluxes at 8 and 12 $~\mu$m, might be particularly useful in distinguishing compact PDRs from \ion{H}{2} regions. PAH emission features around 7-8 $\micron$ and 12 $\micron$ are highly sensitive to the PAH ionization state, with the 7-8 $\micron$ lines becoming much stronger than the 12 $\micron$ lines for ionized PAHs \citep{roe96, dra07}. The PAH ionization fraction in a PDR is related to the compactness of the PDR \citep{roe96}; a more compact PDR will experience a more intense radiation field resulting in a higher PAH ionization fraction, stronger emission around 7-8 $\micron$, and thus a lower log(F12/F8) color. On average, YB PDRs are more compact than typical \ion{H}{2} regions, thus causing the color shift to lower log(F12/F8) colors. KWA15 concluded that a higher fraction of compact objects with a large PAH ionization fraction leads to increasingly negative log(F$_{12}$/F$_8$) values.

Out of 516 YBs in our pilot region, 434, 361, and 351 had 8$~\mu$m, 12 $~\mu$m, and 24 $~\mu$m photometry results with repeated measurements yielding a fractional error of $< 0.5$ and were thus deemed ``reliable" as discussed in \autoref{sec:photo}. We extend the KWA15 analysis to the larger DR2 sample and examine YB colors using additional IR wavelengths. The histogram of the log(F$_{12}$/F$_8$) results for DR2, similar to that shown in KWA15, is shown in \autoref{fig:colorhist}.

KWA15 found that the average value of their sample (-0.19) was more negative than the average color sample of WISE \ion{H}{2} regions (-0.09) \citep{and12} and that the YBs in their sample without RMS counterparts had an even more negative average of -0.23. The colors of our new YB sample yields an even more negative mean log(F$_{12}$/F$_8$) value of $-0.43\pm 0.15$.  Size measurements (based on MWP user radii) confirm that our new YB sample indeed contains a larger number of very compact sources. Whereas the mean ($\pm$ standard deviation) diameter of the RMS-matched YBs with distances from KWA15 is 0.94 pc $\pm$ 0.71 pc, the mean diameter of all YBs with good distance estimates in our current sample is 0.75 pc $\pm$ 0.53 pc. The median diameters are smaller (0.77 pc and 0.63 pc, respectively), as relatively few YBs are significantly larger than 1 pc. The mean diameter of the subset of YBs with counterparts in catalogs of \ion{H}{2} regions and MYSOs (CORNISH, RMS, WISE-C,G,K) is 0.96 pc $\pm$ 0.62 pc and the mean ($\pm$ standard deviation) log(F12/F8) color of these objects is -0.28 $\pm$ 0.31; however, the mean diameter of YBs lacking such counterparts is 0.65 $\pm$ 0.45 pc and the mean log(F12/F8) color of this subset is -0.47 $\pm$ 0.24.
Many compact YBs were missed in the KWA15 sample, which was drawn from images at lower zoom levels, where citizen scientists would have been more likely to notice large and bright YBs. Additionally, only YBs with RMS counterparts had associated distances in the KWA15 sample; our current sample indicates that YBs with counterparts in catalogs of \ion{H}{2} regions and MYSOs are biased to physically larger objects.

In addition to measurements at 8 and 12 $~\mu$m, we use longer MIR wavelengths at 24 and 70 $~\mu$m to examine the YB population in our DR2 sample. We show a MIR color-color plot in \autoref{fig:ccp} for the 219 YBs in our pilot region with reliable 8-$\mu$m and 24-$\mu$m photometry as well as counterparts in the Hi-GAL survey (EMS17). The color-color plot indicates sources with cross-matches in the RMS and WISE catalogs. The average \ion{H}{2} region colors and \ion{H}{2} region - PNe cut-off colors established by \citet{and12} are also indicated in \autoref{fig:ccp}. WISE sources were separated into categories by \citet{and14}:  ``K'' for known, ``G'' for group, ``C'' for candidate, ``Q'' for radio quiet. YBs with counterparts in the RMS catalog and/or WISE K, G, and C sources are, unsurprisingly, grouped near the expected colors for \ion{H}{2} regions. We find a significant number of sources with WISE Q (n = 44) or no counterpart (n = 101). These WISE Q and no-association sources are predominantly clustered in a separate region of the color-color plot than the WISE K, G, C, and RMS sources, with colors brighter at both 8 $\mu$m and 70 $\mu$m than they are at 24 $\mu$m. A possible interpretation of this distribution (similar to that described by \cite{cha95} for 12/25/60 $\micron$ observations of \ion{H}{2} regions) is that the log(F$_{24}$/F$_8$) color acts as a  measure of the hardness of the incident ultraviolet (UV) radiation field, where a hard UV field , with abundant photons beyond the Lyman limit, would destroy PAH 8$~\mu$m emitters, and a softer UV field would excite the PAHs without destroying them \citep{des90, gia94}. Thus a low log(F$_{24}$/F$_8$) value may indicate a region with a soft UV field consistent with environments of intermediate-mass stars. The log(F$_{70}$/F$_{24}$) axis may measure dust temperature, where emission from a cooler environment would peak at longer wavelengths. Consequently the region traced by the WISE Q and no-association YBs is consistent with their being cooler and lower-mass than the YBs in the upper left of \autoref{fig:ccp}.
 
We have shown that on average, YBs with CORNISH, RMS, and WISE-C,G,K matches are larger than those lacking counterparts with either \ion{H}{2} regions or MYSOs ($\sim 1$ pc vs. $\sim 0.7$ pc). They are also more luminous (see \autoref{sec:samps}) and they have higher log(F12/F8) colors ($\sim -0.3$ vs. $\sim -0.5$) and log(F24/F8) colors (see \autoref{fig:ccp}). In addition to distinguishing YBs by mass, is it possible to distinguish evolutionary sequences using MIR colors? 

If we assume that a smaller YB is also younger, we can distinguish the youngest YBs by examining only compact YBs, which we define as those with MWP user diameters less than 0.3 pc. There are only 14 compact YBs that have \ion{H}{2}-region or MYSO matches and reliable colors at 8 and 12 $\micron$. These sources have log(F12/F8) = -0.42 $\pm$ 0.22, more negative than the full massive YB sample. The 47 compact YBs with only WISE-Q or no associations do not show significantly different log(F12/F8) colors than their larger counterparts, with log (F12/F8) = -0.52 $\pm$ 0.24. All of these results support the expectation that MYSOs rapidly evolve to \ion{H}{2} regions (see isochrones in Figure 4). This preliminary analysis is encouraging but limited by small sample size. Photometry results for the full DR2 catalog will enable us to better distinguish the mass and evolutionary stages of YBs.

\subsection{Measuring YB Sizes} \label{sec:size}

Is a radius measured by MWP users an accurate representation of YB size? YBs emit radiation across a broad range of wavelengths, and the different wavelengths may trace different extents of a clump and also be observed with different resolution limits. Thus it is not clear how to define the `size' of a YB. For determining the far-IR fluxes of YBs, we utilized their association with Hi-GAL clumps, whose diameters are based on beam-deconvolved 250-$\mu$m Herschel data (EMS17); however, user-measured MWP sizes of YBs are based on their compact `yellow' appearance in combined 24-8-4.5-$\mu$m Spitzer rgb images. Along with the 2D-Gaussian fits (see \autoref{sec:fits}) we have three independent measures of size. For any given YB, there can be significant variation between the different measures employed to estimate YB size. For example, differences in MWP and Gaussian sizes can be greatly affected by a given object's location within a complex background. 

 \autoref{fig:YBsizes} presents a scatter plot comparing the angular radii of YBs as measured by MWP users to 2D-Gaussian standard deviations. This plot indicates that there is a strong correlation between MWP radii and YB sizes as measured by the 2D-Gaussian fits to the 8- and 24-$\micron$ images. The correlation is even stronger when only well-fit YBs - those that satisfy the 4-point criteria detailed in \autoref{sec:fits} - are considered. Although the error bars on user measurements are large, Pearson's correlation coefficient (r) for the 450 unsaturated YBs plotted in the upper panel is 0.73. The error bars are smaller for the subset of 89 well-fit YBs, and the correlation coefficient is 0.81, indicating that the probability these two parameters are uncorrelated is essentially zero. The slope of the linear fit to the full sample is 1.4 (1.5 to the well-fit subset). 
 This indicates that, on average, sizes measured by MWP users are proportional to YB sizes determined from 2D Gaussian fits to the 8- and 24-$\micron$ emission.

 \autoref{fig:sizecumul} presents normalized histograms of the cumulative angular size distributions for MWP radii, 2D-Gaussian FWHM (2.355$\sigma$), and beam-deconvolved Hi-GAL 250-$\micron$ FWHM. 
 Average angular sizes for each distribution are indicated by dashed vertical lines of the same color as the corresponding histogram. The average user-measured MWP radius (11.9$^{\prime\prime} \pm 4.5^{\prime\prime}$) is comparable to the average 2D-Gaussian FWHM (10.2$^{\prime\prime} \pm 5.2^{\prime\prime}$), while the average beam-deconvolved 250-$\micron$ FWHM is somewhat larger (17.9$^{\prime\prime} \pm 7.3^{\prime\prime}$). The larger size at 250-$\micron$ probably reflects a combination of factors, including lower resolution and the likelihood that compact PDRs are still contained within their birth clumps. We conclude MWP user radii appear to be proportional to 2D Gaussian FWHM, such that trends (i.e. which sources are largest vs. smallest) are consistent across both measures of size.

\section{Conclusions} \label{sec:conclusions}

  We return once more to the question posed by KWA15, ``What are YBs?" Our work supports the main conclusion of KWA15: YBs appear to occupy a transitional phase of massive- and intermediate-mass star formation, when a compact PDR has formed around an embedded protostar and the surrounding interstellar medium remains relatively undisturbed. This paper expands on KWA15 by measuring the physical properties of YBs. We find that a ``typical" YB is sub-parsec in size and has physical properties consistent with being a B-type protostar, but YBs are a heterogeneous sample of objects that span a wide range of mass and evolutionary stage. The percentage of YBs that represents massive vs. intermediate-mass star formation varies with the method used to probe YB properties. The properties of the YBs in our pilot region are summarized below:
\begin{itemize}
    \item 30\% of the YBs have spatial cross-matches with at least one tracer of massive star formation in the CORNISH, RMS, or WISE (C,K,and G sources) catalogs (\autoref{sec:cats}). This is lower than the number of YB cross-matches with tracers of high-mass star formation reported in KWA15, and  reflects the larger number of lower-mass YBs identified in MWP DR2. We expect a significant fraction of these to be precursors to optically-revealed Herbig Ae/Be nebulae. 
    \item YBs span several orders of magnitude in luminosity, and are embedded in dense clumps that likewise span several orders of magnitude in mass. The mass and luminosity ranges are consistent with the clumps being sites of intermediate and massive star formation. The masses are between $0 < \log(M) < 4$ and luminosities between $1 < \log(L) < 5$, with a mean $\log\left(M_\sun\right)$ of 2.37 $M_{\sun}$ and mean $\log\left(L_\sun\right)$ of 3.30. Clumps containing YBs are more massive and more luminous than the protostellar clumps in the full Hi-GAL catalog (\autoref{sec:props}).
    \item Approximately 24\% of clumps containing YBs have $L/M$ consistent with being \ion{H}{2} region candidates (\autoref{sec:props}).
    \item  21\% of YB sources with Hi-GAL counterparts are above the $\Sigma>1$ g cm$^{-2}$ \citet{km08} threshold for high-mass star formation (\autoref{sec:props}). If the $\Sigma>0.3$ g cm$^{-2}$ \citet{lop2010} threshold is used, the proportion increases to 53\% (\autoref{sec:samps}). 
    \item YB bolometric luminosities and bolometric temperatures indicate that the YB sample is slightly more evolved than the full protostellar sample measured by the Hi-GAL catalog (\autoref{sec:props}).
    \item YBs have very heterogeneous structures, especially at shorter MIR wavelengths, and a majority are not well-fit by 2D-Gaussian models (\autoref{sec:fits}). In particular, evidence of multiple sources visible only in the 8-$\mu$m images of YBs indicates many YBs may be compact PDRs encompassing multiple low- and/or high-mass stellar embryos or hyper-compact (HC), ultra-compact (UC), or compact \ion{H}{2} regions \citep{mot18}.
    \item {A large proportion of YB-WISE-Q sources ($\sim 37\%$) are likely associated with massive star-forming regions at a young, radio-quiet, evolutionary stage. The remaining $\sim 63\%$ of YB-WISE-Q sources are associated with regions of intermediate-mass star formation (\autoref{sec:samps}).}
    \item The MIR colors of YBs indicate that many of them are highly compact, with a large PAH ionization fraction. Of the 219 sources in our pilot region with reliable 8 and 24$~\mu m$ photometry and Hi-GAL counterparts, 46\% of the sources have no counterpart in the WISE, CORNISH, or RMS catalogs and 20\% have WISE-Q counterparts. These sources have MIR colors consistent with being lower-mass and cooler sources than the YBs that have CORNISH, RMS, and/or WISE K, G, or C counterparts (\autoref{sec:colors}).
    \item While the size measurement of a YB depends on wavelength and is complicated by resolution limits,  MWP user measurements reasonably trace the extents of YB PDRs (\autoref{sec:size}). MWP user measurements, 2D-Gaussian fits, and distance-rescaled Hi-GAL clump diameters (\autoref{sec:props}), indicate the majority of YBs have sub-parsec diameters and are typically embedded in sub-parsec clumps.

\end{itemize}

The above evidence suggests that $\gtrsim 20\%$ of YBs contain high-mass star formation and could go on to produce expanding \ion{H}{2} regions that produce MIR bubbles, thus we expect $\gtrsim 100$ of the YBs in this region will eventually go on to become bubbles. The accretion phase of a massive protostar should be on the order of $\sim$10$^5$ years, while the lifetime of an expanding \ion{H}{2} region should be on the order of $\sim$10$^6$ years \citep{mot18}. Therefore, we would expect a roughly 10:1 bubbles to massive YBs.
The MWP DR2 catalogs \citep{jay19} identified a $\sim 1:2$ ratio of bubbles to YBs in our pilot region, and this ratio becomes $\sim$ 4:1 when only massive YBs are considered. We note that the bubble catalog of \citet{jay19} has been more rigorously culled of possible false detections, whereas our catalog contains all YBs identified by MWP users, which may partially account for the apparent remaining overcount of YBs compared to bubbles. 

Many of the intermediate-mass YBs do not have existing catalogs of their physical properties and colors. Thus, this YB pilot-region study and future full YB catalog provide a valuable new data-set of properties for intermediate-mass star-forming regions. We will expand the work from this pilot region of 516 YBs to the full catalog of over 6,000 YBs.

\acknowledgements

 The authors thank the numerous MWP volunteers whose efforts made the DR2 dataset possible.  The authors thank our anonymous referee for providing many helpful suggestions for improving this paper. We also thank Davide Elia for a useful discussion of error estimates for quantities in the Hi-GAL compact source catalogue. This publication uses data generated via the Zooniverse.org platform, development of which is funded by generous support, including a Global Impact Award from Google, and by a grant from the Alfred P. Sloan Foundation. This research has made use of the VizieR catalogue access tool, CDS, Strasbourg, France (DOI: 10.26093/cds/vizier); Astropy,\footnote{http://www.astropy.org} a community-developed core Python package for Astronomy \citep{astropy1, astropy2}; data products from the Wide-field Infrared Survey Explorer, which is a joint project of the University of California, Los Angeles, and the Jet Propulsion Laboratory/California Institute of Technology, funded by the National Aeronautics and Space Administration; NASA's Astrophysics Data System Bibliographic Services; and observations made with the Spitzer Space Telescope, which was operated by the Jet Propulsion Laboratory, California Institute of Technology under a contract with NASA. Authors KD, AP, JM, LT, AC, and SS were supported during work on this project by The Murdock Charitable Trust under grant number NS-2016246. GWC was in part supported by a Research Seed Grant through the Illinois Space Grant Consortium. 
 
\software{Astropy \citep{astropy1, astropy2}, Matplotlib \citep{matplotlib}, TOPCAT \citep{topcat}}

\bibliographystyle{aasjournal}

\end{document}